\documentclass[aps,prd,preprintnumbers,twocolumn]{revtex4}
\usepackage{graphicx}
\usepackage{epsfig}
\usepackage{color}
\usepackage{amsmath}
\usepackage{epsf,amssymb,amsfonts,amsmath}
\usepackage{slashed}
\usepackage{float}
\newcommand{\cL}{{\mathrm L}}   \newcommand{\cR}{{\mathrm R}}

\def\p{\partial}

\def\be{\begin{equation}}
\def\ee{\end{equation}}
\def\bea{\begin{eqnarray}}
\def\eea{\end{eqnarray}}

\newcommand{\beq}{\begin{equation}}
\newcommand{\eeq}{\end{equation}}
\newcommand{\beqa}{\begin{eqnarray}}
\newcommand{\eeqa}{\end{eqnarray}}
\newcommand{\beqar}{\begin{eqnarray*}}
\newcommand{\eeqar}{\end{eqnarray*}}

\renewcommand{\eqref}[1]{(\ref{#1})}

\def\p{\varphi}

\catcode`\@=12

\def\p{\partial}

\begin{document}

\title{A holographic model for the baryon octet}
\author{Zhen Fang}
\email{fangzhen@ucas.ac.cn}
\affiliation{School of Physical Sciences, University of Chinese Academy of Sciences, No.19A Yuquan Road, Beijing 100049, China}

\begin{abstract}
By adopting the nonlinear realization of chiral symmetry, a holographic model for the baryon octet is proposed. The mass spectra of the baryon octet and their low-lying excited states are calculated, which show good consistency with experiments. The couplings of pion to nucleons are derived in two gauges and are shown to be equivalent with each other. It also shows that only derivative couplings of pion to nucleons appear in this holographic model. The coupling constant is then calculated.

\end{abstract}

\maketitle

\section{Introduction}

In the recent decades, there have been many studies on AdS/QCD since the discovery of AdS/CFT \cite{Maldacena:1997re,Gubser:1998bc,Witten:1998qj}. Various holographic QCD models have been constructed, either in the top-down approach \cite{Kruczenski:2003uq,Sakai:2004cn,Sakai:2005yt} or in the bottom-up approach \cite{DaRold:2005zs,Erlich:2005qh,Brodsky:2014yha}. Many researches have been focused on the phenomenology of low-energy QCD, such as the hadron spectrum \cite{Gherghetta:2009ac,Sui:2009xe,Li:2012ay,Cui:2016ocl}, the deconfining or chiral phase transition \cite{Herzog:2006ra,Chelabi:2015cwn,Chelabi:2015gpc,Fang:2015ytf,Fang:2016nfj}, and the hadron form factors \cite{Brodsky:2014yha} etc. The AdS/QCD method has shown powerful ability in the description of low-energy QCD phenomenon.

In this paper, we focus on the baryon properties and try to construct a sensible holographic model for the baryon octet, based on the AdS/QCD model for mesons \cite{DaRold:2005zs,Erlich:2005qh}, which have shown superiority in respect of phenomenology. Indeed, considerable efforts have been made in the construction of the holographic baryon models, such as in \cite{deTeramond:2005su,Pomarol:2008aa}. In \cite{Hong:2006ta}, the authors built a holographic model for spin-$\frac{1}{2}$ baryons, which naturally incorporates the important phenomenon of parity-doublet \cite{Jaffe:2006jy}. However, one large limitation is that they only address the parity-doublet of nucleons and their excited states. To generalize the model, one cannot bypass the baryons with heavy quarks, which makes the issue about chiral symmetry troublesome. The meson physics in three-flavor case has been considered in the holographic framework in \cite{Sui:2010ay}. For baryons with three flavors, chiral symmetry is believed to be broken from the beginning and the picture of chiral symmetry breaking in \cite{Hong:2006ta} seems to be invalid.

Note that the holographic construction in \cite{Hong:2006ta} shares similarities with the linear sigma model and the massless modes of baryons before chiral symmetry breaking form the representation of the chiral group. However, for heavy baryons, it is more suitable to accept the nonlinear realization of chiral symmetry in the construction of baryon models \cite{Donoghue:1992dd}. Following this way, a holographic model for the baryon octet will be given. The whole mass spectra of the baryon octet and their low-lying excited states will be calculated, which show good agreement with experiments. we also show that this holographic model only contains gradient couplings of pion with baryon fields, at least in the chiral limit, which is in parallel with the nonlinear construction of the effective baryon model in the 4D field-theoretic method \cite{Weinberg:1966fm}.

\section{Outline of the model}

As in \cite{DaRold:2005zs,Erlich:2005qh}, we take a slice of $\mathrm{AdS}_5$ space-time with the following metric ansatz:
\begin{equation}\label{metric}
ds^2=e^{2A(z)}\left(\eta_{\mu\nu}dx^{\mu}dx^{\nu}-dz^2\right),\quad \epsilon \leqslant z \leqslant z_m ,
\end{equation}
where $A(z)=-\mathrm{ln}\,z$, $\eta_{\mu\nu}=\mathrm{diag}(+,-,-,-)$ and the IR brane $z_m=1/\Lambda_{\mathrm{QCD}}$.

To produce the possible parity-doubling pattern of excited baryon states, we follow \cite{Hong:2006ta} and introduce two bulk baryon fields $B_{1,2}$ which correspond to the chiral baryon operators $\mathcal{O}_{L,R}$ in the boundary respectively, but with the isodoublet in two-flavor case replaced by the baryon octet in three-flavor case:
\begin{equation}\label{Baryon octet}
B = \sum_{a=1}^8 \frac{\lambda^a B^a}{\sqrt{2}} = \left(
\begin{array}{ccc}
\frac{\Sigma^0}{\sqrt{2}}+\frac{\Lambda}{\sqrt{6}}       &          \Sigma^+                                             &        p                        \\
\Sigma^-                                                 &          -\frac{\Sigma^0}{\sqrt{2}}+\frac{\Lambda}{\sqrt{6}}  &        n                        \\
\Xi^-                                                    &                \Xi^0                                          &        -\frac{2\Lambda}{\sqrt{6}}\\
\end{array}
\right).
\end{equation}

The holographic model is formulated on the basis of the chiral gauge group $G=SU(3)_L\times SU(3)_R$, which is broken spontaneously into a vectorial subgroup $H=SU(3)_V$. The bulk action comprises two sectors, one of which containing the covariant kinetic terms and the bulk mass terms can be written as
\begin{equation} \label{baryon-action1}
S_{b1} = \int d^5x \sqrt{g} \sum_{j=1}^2 \mathrm{Tr} \left( \bar{B}_j \,i\, e_A^M\Gamma^A \nabla_M B_j - m_5 \bar{B}_j B_j \right) ,
\end{equation}
where the metric $g_{MN}=e_M^A e_N^B\eta_{AB}$ with the vielbein $e_M^A=\frac{1}{z} \eta_M^A$, the Dirac matrices $\Gamma^A=(\gamma^{\mu},-i\gamma^5)$ satisfy $\left\{\Gamma^A,\Gamma^B\right\}=2\eta^{AB}$, and the mass of the bulk baryon fields $m_5$ can be determined by the AdS/CFT correspondence as $m_5=\pm \frac{5}{2}$ with the sign dictated by the analysis of the chirality of baryon fields (conventionally, we set $m_5=\frac{5}{2}$ for $B_1$ and $m_5=-\frac{5}{2}$ for $B_2$) \cite{Hong:2006ta}. The Lorentz and gauge covariant derivatives of the bulk baryon fields have the form as follows
\begin{eqnarray}
\nabla_{M} B_1 &=& \partial_M B_1 + \frac{i}{4}\omega_M^{AB}\Gamma_{AB}B_1 - i[\mathcal{L}_M,B_1] ,  \\
\nabla_{M} B_2 &=& \partial_M B_2 + \frac{i}{4}\omega_M^{AB}\Gamma_{AB}B_2 - i[\mathcal{R}_M,B_2] ,
\end{eqnarray}
where the spin connection $\omega_M^{AB}= \partial_z A(z)(\delta_M^A\delta_z^B-\delta_z^A\delta_M^B)$, $\Gamma^{AB}=\frac{1}{2i}\left[\Gamma^A,\Gamma^B\right]$ is the Lorentz generator for spinor fields, and $\mathcal{L}_M (\mathcal{R}_M)$ can be formulated by the gauged Maurer-Cartan $1$-forms as
\begin{equation}
\mathcal{L}_M = i\,\xi^{\dagger}D_M \xi, \quad \mathcal{R}_M = i\,\xi D_M \xi^\dagger
\end{equation}
with the covariant derivatives defined by
\begin{eqnarray}
D_M \xi &=& \partial_M \xi-i (A_L)_M \xi, \\
D_M \xi^\dagger &=& \partial_M \xi^{\dagger}-i (A_R)_M\xi^{\dagger},
\end{eqnarray}
where $A_L^M$, $A_R^M$ are the chiral gauge fields and $\xi=e^{i T_a \pi_a}$ is the element of the coset space $G/H$ ($T_a=\lambda_a/2$ is the generator of $SU(3)$ group with $\lambda_a$ the Gell-Mann matrices and $\pi_a$ denotes the bulk pseudoscalar field). We can see that $\mathcal{L}_M$, $\mathcal{R}_M$ contain both the gauge fields and the pseudoscalar fields, and are indeed the only terms containing these fields in the model, which is different from that in \cite{Hong:2006ta}.

Note that in the nonlinear representation the bulk baryon fields $B_j$ only transforms under subgroup $H$ as $B_j^\prime = H B_j H^\dagger$. The element $\xi(\pi)$ of the coset space $G/H$ transforms under the chiral gauge group G as $\xi^{\prime}=L \xi(\pi) H^{\dagger}=H \xi(\pi) R^{\dagger}$ with $L \in SU(3)_L$, $R \in SU(3)_R$, and the chiral gauge fields $A_L^M$ and $A_R^M$ transform in the following way:
\begin{eqnarray}
(A_L)_M^{\prime} &=& L (A_L)_M L^\dagger + i L \partial_M L^\dagger,  \\
(A_R)_M^{\prime} &=& R (A_R)_M R^\dagger + i R \partial_M R^\dagger.
\end{eqnarray}
Then the group transformations of $\mathcal{L}_M$ and $\mathcal{R}_M$ can be obtained as
\begin{eqnarray}
\mathcal{L}_M^{\prime} &=& H \mathcal{L}_M H^\dagger + i H \partial_M H^\dagger,  \\
\mathcal{R}_M^{\prime} &=& H \mathcal{R}_M H^\dagger + i H \partial_M H^\dagger.
\end{eqnarray}

The other sector of the bulk action which generates chiral symmetry breaking can be formulated as
\begin{eqnarray}
S_{b2} &=& -\int d^5x \sqrt{g} \left[c_1\mathrm{Tr}\big( \bar{B}_1 \{\chi_{+},B_2\}\big) + c_2 \mathrm{Tr}\big( \bar{B}_1 [\chi_{+},B_2] \big) \right. \nonumber  \\
 & & \left. \qquad + (c_2 - c_1) \mathrm{Tr}\big(\bar{B}_1 B_2 \big)\mathrm{Tr}\chi_{+} + \mathrm{h.c.}  \right] ,
\end{eqnarray}
where h.c. denotes the Hermitian conjugate, $\chi_+$ is related to the three-flavor generalization of the bulk scalar field $X$ (see also Sec. \ref{coupling}) by
\begin{eqnarray}
\chi_{+} = \frac{1}{2} (\xi^\dagger X \xi^\dagger + \xi X^\dagger \xi) ,
\end{eqnarray}
and $\chi_+$ transforms in the nonlinear representation as $\chi_+^\prime = H \chi_+ H^\dagger$. The bulk scalar field $X$ can be defined as follows
\begin{eqnarray}\label{X defination}
X \equiv \xi \left(X_0+T_0 S_0 + T_a S_a \right) \xi ,
\end{eqnarray}
where $S_0$, $S_a$ represent the scalar singlet and octet, $X_0$ is the vacuum expectation value (VEV) of the scalar field $X$ and is presumed to be of the following form:
\begin{equation}\label{VEV-X}
X_0=\frac{1}{2}\left(
                         \begin{array}{ccc}
                           v_u &     &     \\
                               & v_u &     \\
                               &     & v_s \\
                         \end{array}
                       \right)
\end{equation}
with $v_u=m_u z +\sigma_u z^3$ and $v_s=m_s z +\sigma_s z^3$, where $m_u$, $\sigma_u$ denote the mass and chiral condensate of $u,d$ quarks, and $m_s$, $\sigma_s$ denote the ones of $s$ quark \cite{DaRold:2005zs,Erlich:2005qh}. In terms of Eq.(\ref{X defination}), $\chi_+$ can be reduced to
\begin{eqnarray}
\chi_{+} =  X_0 + T_0 S_0 + T_a S_a ,
\end{eqnarray}
which is only associated with the scalar mesons.

As in \cite{Hong:2006ta}, the bulk baryon fields $B_j$ can be written in the chiral form: $B_j=B_{jL}+B_{jR}$ with $B_{jL,R}$ decomposed by the Kaluza-Klein (KK) and Fourier decomposition as
\begin{equation} \label{fLR}
B_{jL,R}(x,z)=\sum_n\int\frac{d^4p}{(2\pi)^4}e^{-ipx}f_{jL,R}^{(n)}(z)\psi_{L,R}^{(n)}(p) ,
\end{equation}
where $f_{jL,R}$ are the Kaluza-Klein profiles of the bulk baryon fields, $\psi_{L,R}(p)$ are the $4$D spinors satisfying $\gamma^5 \psi_L(p)=\psi_L(p),\,\gamma^5 \psi_R(p)=-\psi_R(p)$ and $\slashed{p}\,\psi_{L,R}(p)=|p|\psi_{R,L}(p)$. Now in terms of the profiles $f_{jL,R}$ (note that the superscript $n$ has been neglected), the equation of motion (EOM) for the baryon octet can be derived from the variation of the bulk action as follows
\begin{eqnarray}\label{EOM}
\begin{pmatrix} \p_z - \frac{9}{2z}  &   -\frac{v(z)}{2z}  \\  -\frac{v(z)}{2z}  &   \p_z+\frac{1}{2z} \end{pmatrix}\,\begin{pmatrix} f_{1L}\\f_{2L}\\ \end{pmatrix}  &=&  -|p|\,\begin{pmatrix} f_{1R}\\f_{2R} \end{pmatrix} , \nonumber \\
\begin{pmatrix} \p_z+\frac{1}{2z} &  \frac{v(z)}{2z}  \\  \frac{v(z)}{2z}   &   \p_z-\frac{9}{2z} \\ \end{pmatrix}\,\begin{pmatrix} f_{1R}\\f_{2R}\\ \end{pmatrix}  &=&  |p|\,\begin{pmatrix} f_{1L}\\f_{2L}\\ \end{pmatrix} ,
\end{eqnarray}
where $v(z)$ has the following form for different particles in the baryon octet:
\begin{equation}\label{VEV}
v(z) =
\begin{cases}
(3\,c_2-c_1)\,v_u(z)  & \text{for\ $(p,n)$,} \\
(2\,c_2-\frac{4\,c_1}{3})\,v_u(z)+(\frac{c_1}{3}+c_2)\,v_s(z)  & \text{for\ $\Lambda$,} \\
2\,c_2\,v_u(z)+(c_2-c_1)\,v_s(z)  & \text{for\ $\Sigma$s,} \\
(c_2-c_1)\,v_u(z)+2\,c_2\,v_s(z)  & \text{for\ $\Xi$s.}
\end{cases}
\end{equation}

From the parity transformations of the baryon fields, we get $f_{1L} = f_{2R}$ and $f_{1R} = -f_{2L}$ for even-parity states, while $f_{1L} = -f_{2R}$ and $f_{1R} = f_{2L}$ for odd-parity states \cite{Hong:2006ta,Fang:2016uer}. As in \cite{Hong:2006ta}, by solving Eqs.(\ref{EOM}) with the boundary conditions $f_{1L}(\epsilon\rightarrow 0)=0$ and $f_{1R}(z_m)=0$, we obtain the mass spectra of the baryon octet and their excited states.

\section{Mass spectra of the baryon octet and their excited states}

To get the baryon spectrum, we first fix the same parameters as the ones in \cite{Hong:2006ta} by fitting the masses of the nucleons $(p,n)$ and their first excited state N(1440): $m_u=0$, $\sigma_u=(198 \,\mathrm{MeV})^3$, $z_m=(205 \,\mathrm{MeV})^{-1}$ and $3\,c_2-c_1=14.8$.
Then we take $m_s=100 \,\mathrm{MeV}$ and $\sigma_s=\sigma_u=(198 \,\mathrm{MeV})^3$, and the last parameter $c_2$ fixed by the mass of $\Lambda$ to be $c_2=4.52$. The experimental and numerical results for the baryon-octet spectrum are shown in Tabel \ref{baryon-spectrum1}, where we can see that the model predictions for the masses of $\Sigma$s and $\Xi$s match with experimental data very well. It can be easily shown that the mass hierarchy of different particles in the baryon octet is exclusively due to the large mass of the strange quark.
\begin{table}[H]
\begin{center}
    \begin{tabular}{|c|c|c|c|c|}
       \hline
      Baryon octet  & $ p,n $ & $\Lambda$ & $\Sigma$s & $\Xi$s   \\
       \hline\hline
       Exp.(MeV) & $939^*$ & $1115^*$ & 1190 & 1320    \\
       \hline
       Model (MeV) & 939 & 1115 & 1184 & 1321   \\
       \hline
     \end{tabular}
\end{center}
\caption{Experimental value and model predictions for the mass spectrum of the baryon octet. * indicates input value, and the experimental data are taken from \cite{Agashe:2014kda}.}
\label{baryon-spectrum1}
\end{table}
\begin{table}[H]
\begin{center}
    \begin{tabular}{|c|c|c|c|c|}
       \hline
      1st even  & $ p,n $ & $\Lambda$ & $\Sigma$s & $\Xi$s  \\
       \hline\hline
       Exp.(MeV) & $1440^*$ & 1600 & 1660 & ---    \\
       \hline
       Model (MeV) & 1440 & 1548 & 1598 & ---   \\
       \hline\hline
     1st odd  & $ p,n $ & $\Lambda$ & $\Sigma$s & $\Xi$s  \\
       \hline\hline
       Exp.(MeV) & 1535 & 1405 & 1620 & ---    \\
       \hline
       Model (MeV) & 1505 & 1588 & 1627 & ---   \\
       \hline
     \end{tabular}
\end{center}
\caption{Experimental value and model predictions for the mass spectra of the first excited even-parity and odd-parity states.}
\label{baryon-spectrum2}
\end{table}
Table \ref{baryon-spectrum2} lists the masses of the first excited baryon-octet states with both even and odd parities, which are also consistent with experiments, except for the first odd-parity state of $\Lambda$. It should be noted that the masses of the excited states of $\Xi$s are not given due to the lack of experimental data. For the masses of higher excited states, this model cannot give consistent results with experiments, the reason of which might be attributed to the sharp IR cutoff of the fifth dimension of $\mathrm{AdS}_5$, which can be remedied through a soft dilaton term \cite{Fang:2016uer,Karch:2006pv}.

It should be remarked that the baryon masses in this model are acquired by a different way from that in \cite{Hong:2006ta}, where the VEV of the scalar field $X$ in a Yukawa coupling term breaks the chiral symmetry and gives the nucleon mass. In our model with a nonlinear realization of chiral symmetry, the chiral symmetry of the baryon action per se has been broken from the beginning, which can be seen from the group transformations of the baryon fields $B_j$ that keep the action invariant. However, it is still the VEV of the scalar field $\chi_+$ that contributes to the mass spectrum of the baryon octet.

A unique feature of this holographic framework is that one cannot separate the explicit chiral symmetry breaking terms from the ones generating spontaneous symmetry breaking in the bulk action as the quark mass and the chiral condensate are sewed into a single term, i.e., the VEV of the bulk scalar field $X_0$. In the three-flavor case, $X_0$ does not vanish due to the large mass of $s$ quark, even if there is no spontaneous chiral symmetry breaking. It leads us to accept the nonlinear realization of chiral symmetry. However, as the bulk baryon action does not realize the full chiral symmetry in the nonlinear representation, the baryon states cannot form the Wigner-Weyl modes of the chiral group, which might cause some vagueness for the explanation of the parity-doublet pattern of excited baryons. We will not go into details about the reason of the parity-doubling phenomenon, which is still inconclusive \cite{Jaffe:2006jy}.

\section{The pion-nucleon coupling} \label{coupling}

In the above, we construct a holographic model for the baryon octet using a nonlinear realization of chiral symmetry, which has been proved more suitable for describing baryons with larger mass \cite{Donoghue:1992dd}. The explanation of the parity-doubling pattern in terms of two baryon fields $B_{1,2}$ with inverse chirality (parity) has been incorporated in this framework, though the bulk baryon action does not realize the full chiral symmetry. Another characteristic of this holographic model is that there are only derivative interactions of pions with baryons, which is also a desired feature in the 4D effective baryon models \cite{Weinberg:1966fm,Donoghue:1992dd}. Next, we will derive the expressions for the pion-nucleon couplings and show directly the nature of derivative interactions.

As the pion wave function is necessary for the derivation of the pion-nucleon coupling, we first present the bulk action of the meson sector which has been studied in \cite{DaRold:2005zs,Erlich:2005qh}:
\begin{equation}\label{meson action}
S_m = \int d^{5}x\,\sqrt{g}\,\mathrm {Tr}\left[|DX|^{2}+3|X|^{2} - \frac{1}{4g_{5}^2}(F_{L}^2+F_{R}^2)\right] ,
\end{equation}
where the covariant derivative of the bulk scalar field $X$ is $D^MX=\p^MX-i A_L^MX+i X A_R^M$ and the chiral gauge field strength is $F_{L,R}^{MN}=\partial^MA_{L,R}^N-\partial^NA_{L,R}^M-i[A_{L,R}^M,A_{L,R}^N]$. The vector and axial-vector fields are defined as $V^M=\frac{1}{2}(A_L^M+A_R^M)$ and $A^M=\frac{1}{2}(A_L^M-A_R^M)$ respectively.

Note that the scalar part of the meson action (\ref{meson action}) can also be recast into a nonlinear form using the Eq.(\ref{X defination}), and the results do not depend on which forms we used. Below we will work in two gauges, as has been used in \cite{DaRold:2005zs} and \cite{Erlich:2005qh} respectively. The results in both gauges will be shown to be equivalent with each other, from which the gauge-independence manifests itself obviously.

Let us first work in the $A_5=0$ gauge accepted in \cite{Erlich:2005qh}. The EOM of the pseudoscalar meson in this gauge can be derived as
\begin{eqnarray}
 \partial_z \left(\frac{\partial_z f_\varphi}{z}\right)+\frac{g_5^2 v_u^2}{z^3}\left(f_\pi-f_\varphi\right) &=& 0  ,    \label{pi-EOM1}\\
 -q^2 \partial_z f_\varphi + \frac{g_5^2 v_u^2}{z^2}\partial_z f_\pi &=& 0  ,      \label{pi-EOM2}
\end{eqnarray}
where $f_\pi(z)$ comes from the KK decomposition of the bulk pion field, and $f_\varphi(z)$ comes from that of the radial component of axial-vector field ($A_\mu^a=A_{\mu\bot}^a+\partial_\mu \varphi^a$). As in the calculation of baryon spectrum, we only consider the $m_u=0$ case in which the pion as Nambu-Goldstone boson has zero mass ($q^2=0$), so we have $f_\pi^\prime(z)=0$, which also indicates with the normalization condition ($f_\pi(z\rightarrow 0)=0$) that $f_\pi(z)=0$. Then Eq.(\ref{pi-EOM1}) can be reduced to the following form:
\begin{eqnarray}\label{pi-EOM11}
z\,\partial_z \left(\frac{\partial_z f_\varphi}{z}\right)-\frac{g_5^2 v_u^2}{z^2}f_\varphi = 0 .
\end{eqnarray}

The pion-nucleon couplings are supplied by the covariant kinetic term of the bulk baryon action (\ref{baryon-action1}), which is the unique term containing the pseudoscalar and gauge fields, as noted above. In the $A_5=0$ gauge, the interaction terms (Lagrange density) of pion and nucleons can be extracted from the action (\ref{baryon-action1}) as
\begin{eqnarray}\label{pi-N-gauge1}
\mathcal{L}_{\pi NN}^{(1)}  &=& z \left[ \left(f_{\pi}-f_{\varphi}\right)\left(|f_{1\cL}|^2-|f_{2\cL}|^2\right)\,\bar{\psi}\gamma^5 \slashed{\partial} \pi^a t^a \psi \right.     \nonumber  \\
& &  \left. + \partial_z f_{\pi} \left(f_{1\cL}^*f_{1\cR}-f_{2\cL}^*f_{2\cR}\right)\,i\,\bar{\psi}\gamma^5 \pi^a t^a \psi\right]   \nonumber  \\
&=& -z f_{\varphi}\left(|f_{1\cL}|^2-|f_{2\cL}|^2\right)\,\bar{\psi}\gamma^5 \slashed{\partial} \pi^a t^a \psi ,
\end{eqnarray}
where the superscript * denotes complex conjugate, $t^a=\frac{\sigma^a}{2}$ with $\sigma^a$ the Pauli matrices and $\psi$ denotes the 4D isodoublet of nucleons. The Lagrange density (\ref{pi-N-gauge1}) shows obviously the derivative coupling of pion to nucleons, which is a characteristic feature of the nonlinear representation. Note that it is different from the holographic baryon model proposed in \cite{Hong:2006ta}, where a Yukawa coupling term was introduced for chiral symmetry breaking, which mimics the structure of the linear sigma model.

For comparison, we take another gauge-fixing scheme used in \cite{DaRold:2005zs} to work out the coupling of pion to nucleons, and show that the result is equal to the one in the $A_5=0$ gauge. The pion meson in this gauge is associated with the axial-vector and pseudoscalar fields which can be extracted (up to quadratic terms) from the action (\ref{meson action}) as follows:
\begin{eqnarray}\label{pi-A-action}
S_{\pi A} &=& \int d^{5}x\,\bigg[-\frac{1}{2 g_{5}^2 z}\left(\partial_\mu A^a_\nu \partial^\mu A^{a\nu}-\partial_\mu A^a_\nu \partial^\nu A^{a\mu} \right.   \nonumber  \\
& & \left.-(\partial_\mu A^a_5-\partial_z A^a_\mu)^2\right)+\frac{v_u^2}{2z^3}\left((\partial_\mu \pi^a-A_\mu^a)^{2}  \right.  \nonumber  \\
& & \left. -(\partial_z \pi^a-A_5^a)^{2}\right)\bigg].
\end{eqnarray}

To eliminate the mixing terms between $A_5^a$ ($\pi^a$) and $A_\mu^a$, we add the following gauge-fixing term:
\begin{equation}
S_{\mathrm{g.f.}} = \int \frac{-d^{5}x}{2\xi_A g_5^2 z}\left[\partial^\mu A_\mu^a-\xi_A z \partial_z \left(\frac{A_5^a}{z}\right)
 +\frac{\xi_A g_5^2 v_u^2}{z^2}\pi^a\right]^2 .
\end{equation}
In the unitary gauge, we take $\xi_A\rightarrow \infty $ to decouple the term $z \partial_z \left(\frac{A_5^a}{z}\right)-\frac{g_5^2 v_u^2}{z^2}\pi^a$ from the $A_\mu^a$ terms. Note that the orthogonal combination of $A_5^a$ and $\pi^a$ remains massless if
\begin{eqnarray}\label{A-pi-relation}
z\,\partial_z \left(\frac{A_5^a}{z}\right)-\frac{g_5^2 v_u^2}{z^2}\pi^a=0.
\end{eqnarray}
With the ansatz $A_5^a(x,z)=f_0(z) \pi^a(x)$, Eq.(\ref{A-pi-relation}) becomes
\begin{eqnarray}\label{A-pi-relation2}
z\,\partial_z \left(\frac{f_0(z)}{z}\right)-\frac{g_5^2 v_u^2}{z^2}f_\pi(z)=0.
\end{eqnarray}

After fixing the gauge, the pseudoscalar part of the action (\ref{pi-A-action}) can be written as
\begin{eqnarray}\label{piaction}
S_{\pi} &=&  \int d^{5}x\,\left[\frac{1}{2 g_{5}^2 z}\partial_\mu A^a_5 \partial^\mu A^a_5 + \frac{v_u^2}{2z^3}\partial_\mu \pi^a \partial^\mu \pi^a \right. \nonumber \\
& & \left. -\frac{v_u^2}{2z^3}(\partial_z \pi^a-A_5^a)^{2}\right],
\end{eqnarray}
from which one can see that the condition for pion to be massless is $\partial_z \pi^a-A_5^a=0$, which implies $f_0(z)=f_{\pi}^\prime(z)$. In this gauge, we can also derive the coupling of pion to nucleons as
\begin{eqnarray}\label{pi-N-gauge2}
\mathcal{L}_{\pi NN}^{(2)}  &=& z \left[ f_{\pi}\left(|f_{1\cL}|^2-|f_{2\cL}|^2\right)\,\bar{\psi}\gamma^5 \slashed{\partial} \pi^a t^a \psi \right.   \nonumber  \\
& &  \left. + \left( \partial_z f_{\pi} - f_0 \right) \left(f_{1\cL}^*f_{1\cR}-f_{2\cL}^*f_{2\cR}\right) \,i\,\bar{\psi}\gamma^5 \pi^a t^a \psi\right]
\nonumber \\
&=& z f_{\pi}\left(|f_{1\cL}|^2-|f_{2\cL}|^2\right)\,\bar{\psi}\gamma^5 \slashed{\partial} \pi^a t^a \psi .
\end{eqnarray}

To prove the equivalence of $\mathcal{L}_{\pi NN}$ in Eqs.(\ref{pi-N-gauge1}) and (\ref{pi-N-gauge2}), we note that Eq.(\ref{A-pi-relation2}) is just the same as Eq.(\ref{pi-EOM11}) in view of the relation $f_0(z)=f_{\pi}^\prime(z)$ derived in the unitary gauge. It indicates that the couplings of pion with nucleons are equivalent with each other in the two gauges. The gradient coupling constant can be defined as
\begin{eqnarray}
\frac{g_A}{F_\pi} \equiv \int_0^{z_m} \frac{dz}{z^4} f_{\pi}(z)\left(|f_{1\cL}(z)|^2-|f_{2\cL}(z)|^2\right),
\end{eqnarray}
where $F_\pi$ denotes the pion decay constant, $f_{1\cL}(z)$ and $f_{2\cL}(z)$ are normalized by the condition:
\begin{equation}
\int_0^{z_m} \frac{dz}{z^4}\left(|f_{1\cL}(z)|^2+|f_{2\cL}(z)|^2\right) = 1 ,
\end{equation}
and $f_{\pi}(z)$ is solved from Eq.(\ref{A-pi-relation2}) with the normalization of $f_0(z)$:
\begin{equation}\label{normalization}
\int_0^{z_m} dz\,\left[\frac{1}{g_{5}^2 z}f_0^2 + \frac{z^3}{g_5^4 v^2}\left(\partial_z\left(\frac{f_0}{z}\right)\right)^2 \right] = 1.
\end{equation}

The pion decay constant $F_\pi$ can be computed from the meson action (\ref{meson action}) as $F_\pi^2 = -(g_5^2 z)^{-1}\partial_z A(0,z)|_{z=\epsilon}$ with $A(q,z)$ the bulk-to-boundary propagator of the axial-vector field \cite{Erlich:2005qh}. Using the parameters fixed by the baryon spectrum, we get $F_\pi\simeq 54.5 \mathrm{MeV}$ and $g_A \simeq 0.33$, which are too small compared with the experimental value: $F_\pi\simeq 92.4 \mathrm{MeV}$ and $g_A \simeq 1.27$ \cite{Agashe:2014kda}.

The discrepancy between the holographic model for spin-$\frac{1}{2}$ baryons and the one for mesons has been revealed in \cite{Hong:2006ta}, where the masses of $N(1440)$ and $N(1535)$ cannot be obtained from the fitting of the meson sectors \cite{DaRold:2005zs}, for which they argued by the possibility of the non-vanishing anomalous dimension of baryons. Another important reason for this inconsistency, as has been noted above, is that the sharp IR cutoff of the $\mathrm{AdS}_5$ metric causes no linear confinement which is manifest in the hadron spectrum \cite{Fang:2016uer,Karch:2006pv}.

The coupling of pion to nucleons may also receive contributions from non-minimal gauge interactions, such as the following magnetic gauge coupling term \cite{Ahn:2009px}:
\begin{eqnarray}
\mathcal{L} &=& i\,\kappa \left[\bar{B}_1 \Gamma^{MN} (\mathcal{F}_L)_{MN} B_1 - \bar{B}_2 \Gamma^{MN} (\mathcal{F}_R)_{MN} B_2 \right],      \nonumber  \\
\end{eqnarray}
where the transformed gauge field strengths are $\mathcal{F}_L^{MN}=\xi^{\dagger}F_L^{MN}\xi$ and $\mathcal{F}_R^{MN}=\xi F_R^{MN}\xi^{\dagger}$. Note that this term also only generates gradient coupling of pion to nucleons.

\section{Summary and Conclusion}

In this paper, we propose a holographic model for the baryon octet with a nonlinear realization of chiral symmetry on the basis of the AdS/QCD model for mesons \cite{DaRold:2005zs,Erlich:2005qh}. The mass spectra of the baryon octet and their low-lying excited states have been calculated, which show good agreement with experiments, especially for the ground states. However, the model cannot give consistent results with experiments for highly excited baryon states, which might be attributed to the sharp IR cutoff of the $\mathrm{AdS}_5$ metric \cite{Fang:2016uer,Karch:2006pv}. We have also shown that the model only contains gradient coupling of pion to nucleons, as in the 4D effective baryon models with nonlinear realization of chiral symmetry \cite{Donoghue:1992dd}. However, the derivative coupling constant is much smaller than that obtained from experimental measurements, which might be rooted in the ignorance of other gauge interaction terms contributing to the coupling of pion to nucleons, as has been noted above.

Note that two bulk baryon fields are introduced in this holographic model, although they transform in the same way under the gauge group $H$. This is an unusual point of the model which is different from the previous studies \cite{Hong:2006ta}. As the bulk action does not realize the full chiral symmetry, the parity-doubling property of excited baryon states cannot be explained with a Wigner-Weyl realization of the chiral group, so other reasons must be called for \cite{Jaffe:2006jy}. On the other hand, this holographic baryon model with nonlinear realization can be easily transformed into a linear representation with the full chiral symmetry restored and the baryon spectrum unchanged. However, it will lead to lengthier action terms and the derivative coupling of pion with baryons will be lost.

Anyway, the first holographic model for the baryon octet has been constructed, which incorporates the parity-doublet pattern of excited baryon states in an unusual way and leads to unique derivative coupling of pion to nucleons. Nevertheless, there is much room to improve the model, and much more low-energy hadron physics related to the baryon octet can be studied in this holographic framework. How to reconcile the physics of the baryon sector and the meson sector in a desired way still needs to be considered.

\vskip 0.5cm
\section{Acknowledgement}
\vskip 0.2cm
The author thanks Doctor Danning Li and Professor Yue-Liang Wu for valuable discussions and generous help, and also thanks Professor M. Traini for useful communications. In addition, the author is also grateful to Professor Cong-Feng Qiao for the application of postdoctoral position.

\end{document}